# Regenerative and Adaptive schemes Based on Network Coding for Wireless Relay Network

Ahmed Hassan M. Hassan, Bin Dai and Benxiong Huang

Department of Electronics and Information Engineering
Huazhong University of Science and Technology
Wuhan 430074, P.R. China
{engahmed5577, nease.dai}@gmail.com and bxhuang@hust.edu.cn

## ABSTRACT

*Recent technological advances in wireless communications offer new opportunities and challenges for relay network. To enhance system performance, Demodulate-Network Coding (Dm-NC) scheme has been examined at relay node; it works directly to De-map the received signals and after that forward the mixture to the destination. Simulation analysis has been proven that the performance of Dm-NC has superiority over analog-NC. In addition, the Quantize-Decode-NC scheme (QDF-NC) has been introduced. The presented simulation results clearly provide that the QDF-NC perform better than analog-NC. The toggle between analog-NC and QDF-NC is simulated in order to investigate delay and power consumption reduction at relay node.*

## 1  INTRODUCTION

Cooperative communication is a specific area of wireless communication that has been extensively explored within the last decade. Ever since the relay network has been appeared, it has led to introduce new challenges, i.e., low reliability, limited throughput, high data rate, etc. Meanwhile, the Network Coding (NC) approach established to be an efficient technique to address same challenges. This pattern includes coding and retransmission of messages at the intermediate nodes of the network. Motivated by these problems, this study deals with an alternative design of fresh forwarding schemes.

Figure 1 depicted the classification of the wireless relay schemes. Ultimately, the wireless relay protocols can be divided into categories [1], [2]: fixed and adaptive schemes. A fixed scheme has two branches known as transparent and regenerative schemes. In transparent schemes, the relay node applies a very simple mathematical operation i.e., multiplication or phase rotation. The significant protocol fall under this category is AF-protocol, etc. In regenerative schemes, the relay amends the received signals and forwards the newer version to its neighbor(s) or destination(s). The few enlisted protocols belonging to this associate are Decode-and-Forward, Detect-and-Forward (DtF), Quantize-and-Forward (QF), etc. Regards to the adaptive scheme, *the selection of suitable protocol is questionable.* Depend on the channel state information (CSI) in the source node and partner, the adaptive scheme classified into two categories: selective and incremental scheme. The difference between the two schemes is that, the early scheme depends on the channel measurement between the cooperation terminals, while another scheme, the feedback message is required from the destination and/or the relay to the source [1].

DOI : 10.5121/ijcnc.2012.4305                                                                                                                                        69

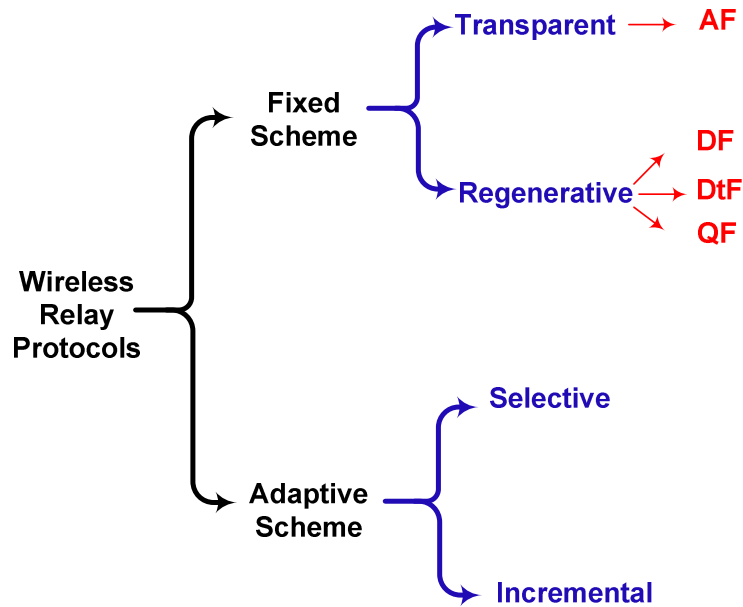

Figure 1 Classification of the wireless relay protocols

In this article, we introduce a new forwarding strategy based on NC to improve the system performance and reduce the system complexity. This strategy known as Quantize Decode-and-Forward based NC (QDF-NC). To the best of our awareness, the QDF-NC is not available in the literature. The ultimate goal of this scheme is to compute the scalar quantization value based on the quantization levels and set the values. By the other phrase, compute scalar quantization value based on quantization levels, prescribed values and finally perform NC. More formally, the relay occupies a quantizer with quantization rule then applies NC to combine the two signals. The different between QDF-NC scheme and Compress-and-Forward (CF) is that the CF-protocol is consisting of a quantizer and index encoder. The quantizer translates the analog signals into digital signals, which are processed by succeeding index encoder to get further compression. In the process of quantization actually we have been mapping the input variable's space to quantize space. We compare the results of the novel scheme with analog-NC. The simulation results approved that our scheme outperformed analog-NC over a wide SNR rang. Furthermore, we investigate an adaptive scheme to allow the relay node to switch between analog-NC, and QDF-NC depends on the hard estimate. The selection of the best scheme is attaining by one source $S_1$ or $S_2$. The main contributions of this article are two-fold:

i. Introduce new strategies at relay node based on NC know as DmNC and QDF-NC.
ii. Investigate flexible protocol at relay node (allow the relay nodes to switch between the best protocol, i.e. analog-NC or QDF-NC). The feasibility of adaptive strategy is mainly demonstrated in this article.

The rest of this work is organized as follows: The related work has presented in Section-2. Section-3 introduces the system models. The strategies and adaptive scheme that relay exploits for forwarding are presented in Section-4. The simulation results and the conclusions are provided in Section-5 and Section-6 respectively.

## 2 RELATED WORK

This section recalls some of the relevant literature in this area. The DtF scheme has been investigated in [3]. In this scheme, the relay node demodulates the received signal and then forwards detected information (modulated information) to the end user(s) (i.e. ***Demodulate-modulate-and-forward***). *Mustapha et al.* in [3] investigated and analyzed the DtF scheme based on the source-relay channel with a bit-interleaved coded modulation structure of the cooperative systems over fading channels. The main objective of this scheme was to provide a destination with the average inter-user channel SNR by sending by the relay. Finally, the destination performs soft input, soft output maximum a posteriori algorithm with limit iterations. In contrary, we investigate new scheme based on NC know as demodulate, NC, modulate and forward (DmNC) scheme. The main function of this scheme is to demodulate the received signal, perform NC then forward baseband signal to the destination. The simulation result shown that the propose scheme is outperforms analog-NC scheme.

Recently, QF scheme has appeared to be an interesting area for a researcher [4], [5], [6], [7]. QF is used instead of sending a new version of a baseband signal after detecting the phase and performs a uniform quantization by the relay [8]. This scheme has to quantize the phase of the received information without any knowledge of source-relay channel quality. In [9] have proposed scheme known as quantized maps-and-forward (QMF) to quantizes the received signal at the noise level and generates a new version of mapping signal before forwarding to destination. In [10], have proposed the joint of QMF and decoding problem as a sum-product algorithm over a factor graph. Compute-quantize scheme with low complexity architecture for distributed antenna has been proposed in [11]. From the previous work, we conclude that the available QF works associate with cooperative communication and there is no work joint NC.

## 3 SYSTEM MODEL

We consider the model system shown in Figure 2 so-called Multiple Access Relay Channel (MARC), which consists of two source nodes, $S_1$ and $S_2$, relay node, $R$, and destination node, $D$. In this network, the two sources wishes to multicast information after un-code/encoded and modulation over AWGN to destination $D$. The key feature of this network is that it contains a relay node $R$.

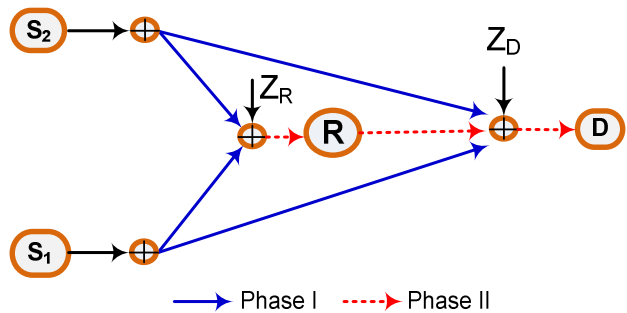

Figure 2 Multiple access channel system model

For application, a very low latency on the data and considerable bit error rate (BER) performance criterion requires. These lead to authorize the convolutional codes in the first choice. We choose the convolutional and BPSK to represent the encoding and modulation technique in case of QDF-NC scheme while the un-code QAM signal to represent DmF-NC.

International Journal of Computer Networks & Communications (IJCNC) Vol.4, No.3, May 2012In first-time of transmission, the two sources emit two messages, $b_{S1}$ and $b_{S2}$, the received information at relay node written as:

$$y_R = \underbrace{\sqrt{P_{S_1}} h_{S_1,R} b_{S_1}}_{y_{S_1}} + \underbrace{\sqrt{P_{S_2}} h_{S_2,R} b_{S_2}}_{y_{S_2}} + Z_R \quad (1)$$

where $b_{S1}$ and $b_{S2}$ are the encoded information, $h_{Si,R}$, $i=1,2$ denotes channel coefficients and the power constraint ($P_{Si}$) at the two sources is $1/(E[\|X_{S_i}\|^2]) \leq P_{S_i}$. For simplicity we assume that $h_{Si,R}$ & $P_{Si}$ are unity.

In our simulation, we used BPSK in case of QDF-NC scheme and QAM to mapping the channel symbols to signal levels [12] in DmF-NC scheme. The BPSK signal and QAM symbols (0, 1, 2, 3) obtain by using equation (2) and (3) respectively:

$$2b_{S_i} - 1 \quad (2)$$

$$(2b_{1,2i+1}|_{odd} + b_{1,2i}|_{even}) \quad (3)$$

Each QAM symbol is consisting of 2 bits (00, 01, 10, 11), these bits are represented in quadrature form as (-1+j, -1-j, 1+j, 1-j) "constellations". Thus, equation (2) generate QAM symbol as:

```
00 → 2*0 + 1*0 = 0
01 → 2*0 + 1*1 = 1
10 → 2*1 + 1*0 = 2
11 → 2*1 + 1*1 = 3
```

In phase two, the relay nodes uses, the analog or the digital processing to transmit the newer version of the signal. The output of the relay is modeled as:

$$y_D^2 = \sqrt{P_R} h_{R,D} f(y_R) + Z_D^2 \quad (4)$$

where $f(.)$ denotes the process applied in the relay node. Here, $Z_{S,D}$ and $Z_{S,R}$ are the additive noise, $h_{S,R}$ and $h_{S,D}$ are the channel coefficients from $S$ to $R$ and $S$ to $D$, respectively, and $h$ is a complex channel coefficient $h = a(t)e^{-jq(t)}$, and $\alpha(t)$ is Rayleigh's distribution and $q(t)$ is the uniform distribution between intervals $-p,p$. They are modeled as $h_c \sim N(0, s_c^2)$ where, $c, s_x^2$ denote the channel link and exponential distribution parameter. This consists of two multiplication components; path loss and fading with zero mean and variance ($\sigma^2$) for each source. The $P_i$ is transmission power at $S$ and $R$ ($i=S, R,$ respectively) and $Z$ is AWGN.

In case of analog-NC, in the second-time slot of transmission the relay amplifies the combination of the received signal then broadcasts. The relay will normalize the received signal by a factor $\beta = \sqrt{E[|y_R|^2]}$, so that the average energy is unity.

## 4 REGENERATIVE AND ADAPTIVE SCHEMES

In this section, we provide the mechanism of schemes that can perform at relay node to forward a new version of received information. These methods so-called DmNC, QDF-NC and adaptive scheme presented as follows:

### 4.1 Demodulate, NC, modulate and forward scheme

The DmNC scheme introduced to improve the system performance. DmNC scheme defines as *directly De-maps the received signals, perform combination (NC) and re-map the combination* to a signal for relay. Specifically, we can write the output of relay node in case of two input signals as in Equ (2). We restrict to un-coded QAM signal to the represent the results of this scheme. We consider MARC as the system model. The process in the relay node depicted in Figure 3 below:

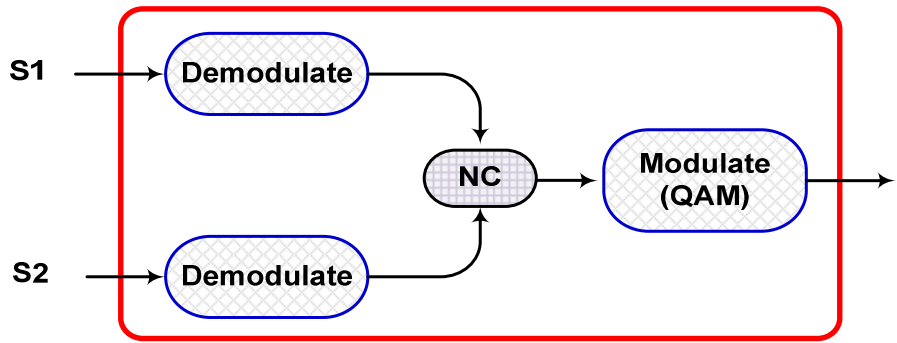

Figure 3 The process of Dm-NC scheme in the relay node

The two sources emit modulate signal (QAM) over AWGN to the relay node simultaneously. The output of the relay send to destination over AWGN after modulates the combination of the received signals. At the end of the receiver, the destination demodulates three signals, two signals from the direct link and one from relay.

The simulation results prove that the Dm-NC outperformed analog-NC over AWGN with un-code QAM signal.

### 4.2 Quantize decode-and-forward based NC scheme

To reduce the complexity of the existing schemes, it is important to introduce a new forwarding scheme, identify as quantize-decode-and-forward based NC (QDF-NC). The processing inside the relay node is depicted in Figure 4 below:

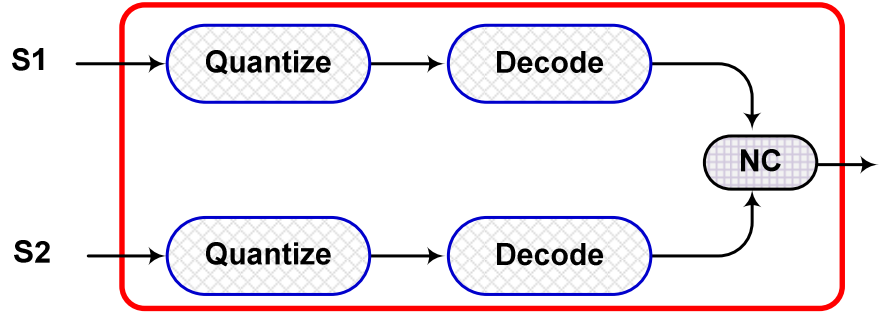

Figure 4 QDF-NC scheme.

The two sources want to send encoded message to the destination through relay node in absent of a direct link. The two sources generating random binary data at rate $R = N/T_s$, where $N$ denotes the string that represents a data packet and $T_s$ is the duration of time slot in which one packet can be transmitted. The relay performs quantizer, decode and then combine the two signals.

The simulation result provides the compression between analog-NC, DF-NC and QDF-NC over AWGN and fading channel. Therefore, QDF-NC scheme outperforms the other two schemes.

### 4.3 Adaptive schemes

The system model investigated in this work is based on [13] and depicted in Figure 5. We apply the system model in MARC model. The whole process from the two sources $S_1$ and $S_2$ to the destination D is described in Figure 5.

At relay node, the LLR is obtaining based on hard decision, then performs QDF-NC or analog-NC based on the probability of BER. The means and variance sample of an equivalent noise process respectively given as:

$$\mu_{Z_R} = \frac{1}{N}\sum_{n=1}^{N}\left(1 - b_R b_R^h\right) \qquad (5)$$

$$\sigma_{Z_R}^2 = \frac{1}{N}\sum_{n=1}^{N}\left(1 - b_R b_R^h - \mu_{Z_R}\right)^2 \qquad (6)$$

where, $b_R^h$ denotes the hard estimates. At the end of a receiver, the destination retrieves the information send by the two sources through direct links in first phase of transmission and from the relay node in the second phase of transmission. The destination performs XoR operation of data received from the relay with decoding the two received messages from the direct links.

Recently, the adaptive schemes introduced to allow the relay node(s) to decide which protocol can be employed to forward the received information depend on the SNR of S→R channel [1], [14]. In this work we allow the relay to make a decision depend on the error probability of the received information [13] i.e. encoding data. The adaptive scheme is described in next Theorem and depicted in Figure 5.

**Theorem:** *The employed protocol will select it depends on: If the error probability of the received information's greater than a given threshold $P_{th}$, then the relay forwards the QDF-NC-scheme. Otherwise, the relay forwards the analog-NC.*

The error probability can be written as [15]:

$$P(BER) = \frac{1}{N} \sum_{n=1}^{N} sign(y_R) \qquad (7)$$

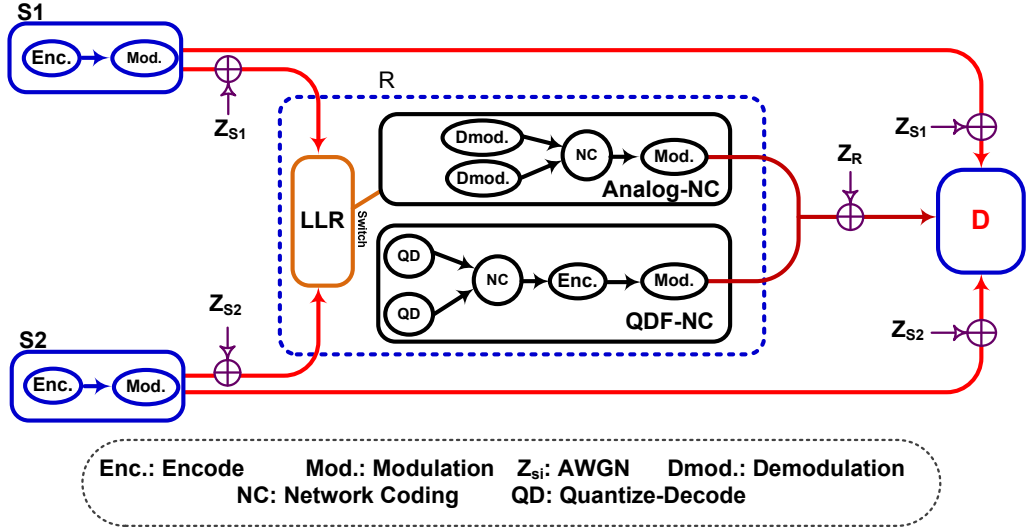

Figure 5 System model of selected QDF-NC or Analog-NC

Depend on the channel link between source(s) and destination, the adaptive scheme is more suitable to forward the information to D. instead of bit error probability (P(BER)), the relay decide which scheme will used to forwards the information. If the value of the P(BER) is greater than a given threshold $P_{th}$, then the relay will use QDF; otherwise analog-NC is performed.

"Hence, this strategy offers a quality-oriented approach to a trade-off between P(BER) and decoder usage (power consumption and implying delay). All nodes take the decision about QDF or analog-NC independent of any other node, only dependent on the local parameter $P_{th}$. The selection of the value $P_{th}$ of depends on the channel code, modulation scheme, channel, energy consumption, network setup, and QoS requirements. It could be adapted according to the number of repeat requests by the following relay. Therefore, the complexity, computational as well as network control, does not increase when the network scales up [13]."

## 5 NUMERICAL RESULTS

In this section, we provide simulations results for regenerative and adaptive schemes. We perform a computer simulation to illustrate the above analysis. In all simulations, we assume that the number of bits per packet is $10^3$, constraint length is 6 trellis matrix is [23 35] and the SRFC Doppler shift is 100. For fair comparison, we present BER curves as functions SNR [dB]. All complex random value of channel coefficients is modeled as AWGN and have $E\{|h_{S_i,R}|\} = E\{|h_{S_i,D}|\} = E\{|h_{R,D}|\} = 1$.

International Journal of Computer Networks & Communications (IJCNC) Vol.4, No.3, May 2012

Figure 6 provides the comparison between analog-NC and DmNC. The simulation results show that the DmNC outperform analog-NC in a wide range of SNR. We can observe from Figure 7 that the QDF-NC scheme outperforms the other two schemes.

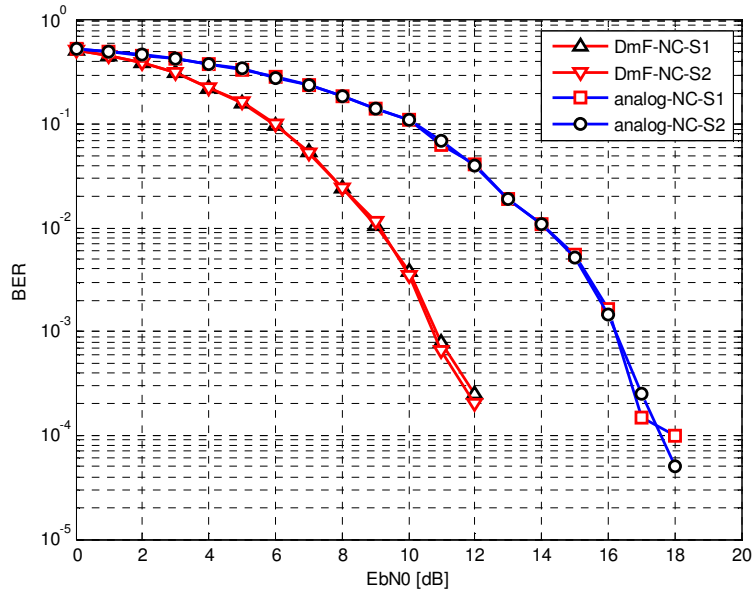

Figure 6 Comparison between DmNC and analog-NC over AWGN with un-code QAM signal

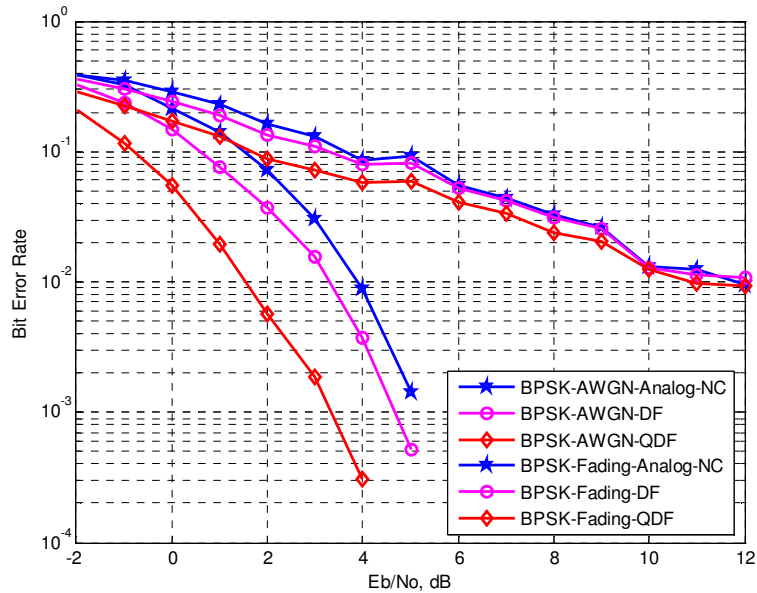

Figure 7 Comparison between analog-NC, DF-NC and QDF-NC schemes

The results of Figure 8 provide the compression between the analog-NC and QDF-NC. We easily observe that the QDF-NC has superior performance than analog-NC over a wide range of SNR. Furthermore, the results of adaptive scheme are presented. The different values 0.2, 0.3 and 0.4 are selected to be $P_{th}$. We can observe that these points represent the point of leaves

QDF-NC curve to analog-NC. If the $P_{th}$ is higher the relay select analog-NC at low SNR and versa vise.

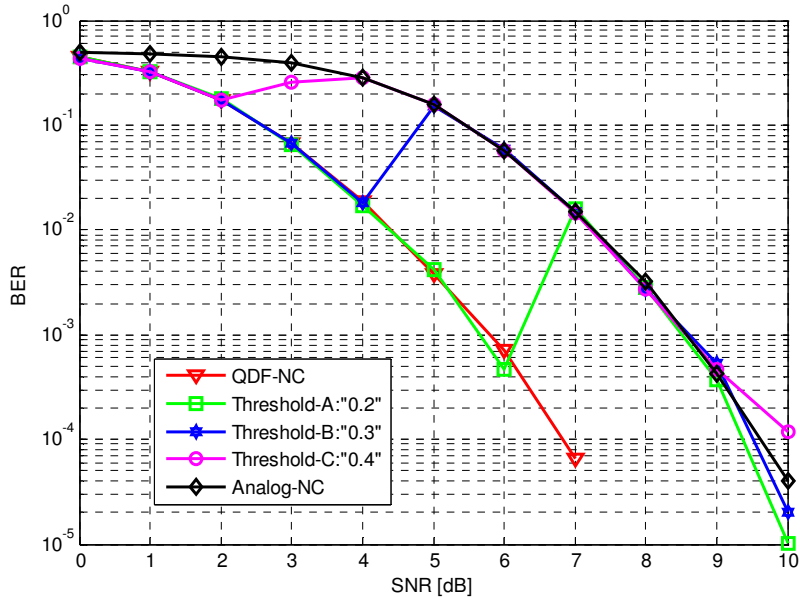

Figure 8 The performance result of analog-NC, QDF-NC and adaptive scheme

For future research the error propagation between the source and destination can be investigate to advance the system model. Furthermore, apply the soft information to the received information need it for reliable communication in case of multiple relays.

## 6   CONCLUSION

In this paper, we have studied the difference between the analog-NC and DmNC scheme with two sources over AWGN. The simulation analysis has demonstrated the improvement in performance of the DmNC scheme over analog-NC scheme. Secondly, we have proposed the QDF-NC scheme compare with analog-NC. The QDF-NC achieved better performance than analog-NC. Furthermore, the adaptive scheme to switch between QDF-NC and analog-NC with MARC over AWGN provided. From the results, we observed that the adaptive scheme useful for ad-hoc network due to limitation of the energy in a node. LDPC codes compare favorably, if we wish to focus on a low SNR or to obtained frame error rate.

The work investigated in this article opens many issues for a future study. The error propagation between the source and destination can investigate to advance the adaptive scheme. Furthermore, apply the soft information to the received information need it for reliable communication in case of multiple relays [16], [17]. The tradeoff between power consumption, complexity and hardware cost will also be considered in our developed model.

## ACKNOWLEDGMENT

This work was supported by the National Natural Science Foundation of China under Grant No.60803005 and the Important National Science and Technology Specific Projects 2009ZX03004-004 and 2010ZX03003-003. Bin Dai is corresponding author.

International Journal of Computer Networks & Communications (IJCNC) Vol.4, No.3, May 2012# REFERENCES

[1] J. N. Laneman, D. N. C. Tse, and G. W. Wornell, "Cooperative diversity in wireless networks: Efficient protocols and outage behavior," *Information Theory, IEEE Transactions on,* vol. 50, pp. 3062-3080, 2004.

[2] M. Dohler and Y. Li, *Cooperative Communications: Hardware, Channel and PHY*: Wiley, 2010.

[3] Mustapha Benjillali and L. Szczecinski, "A simple detect-and-forward scheme in fading channels," *Communications Letters, IEEE,* vol. 13, pp. 309-311, 2009.

[4] B. Djeumou, S. Lasaulce, and A. Klein, "Practical quantize-and-forward schemes for the frequency division relay channel," *EURASIP Journal on Wireless Communications and Networking,* vol. 2007, p. 2, 2007.

[5] I. Avram, N. Aerts, and M. moeneclay, "A novel quantize-and-forward cooperative system: channel parameter estimation techniques," 2010, pp. 1-8.

[6] S. Schwandter and G. Matz, "A practical forwarding scheme for wireless relay channels based on the quantization of log-likelihood ratios," in *International Conference on Signal Processing (ICASSP)*, Dallas, TX 2010, pp. 2502-2505.

[7] A. Chakrabarti, A. Sabharwal, and B. Aazhang, "Practical Quantizer Design for Half-Duplex Estimate-and-Forward Relaying," *IEEE Transactions on Communications,* vol. 59, pp. 74-83, 2011.

[8] M. R. Souryal and H. You, "Quantize-and-forward relaying with M-ary phase shift keying," in *Proceedings of the IEEE Wireless Communications and Networking Conference (WCNC)* Las Vegas, Nev, USA, 2008, pp. 42-47.

[9] A. S. Avestimehr, S. N. Diggavi, and D. N. C. Tse, "Wireless Network Information Flow: A Deterministic Approach," *IEEE Transactions on Information Theory,* vol. 57, pp. 1872-1905, 2011.

[10] V. Nagpal, I. H. Wang, M. Jorgovanovic, D. Tse, Nikolic, x, and B., "Quantize-map-and-forward relaying: Coding and system design," in *48th Annual Allerton Conference on Communication, Control, and Computing (Allerton)*, 2010, pp. 443-450.

[11] S. N. Hong and G. Caire, "Quantized Compute and Forward: A Low-Complexity Architecture for Distributed Antenna Systems," in *IEEE Information Theory Workshop*, Paraty, Brazil, 2011.

[12] K. N. Pappi, A. S. Lioumpas, and G. K. Karagiannidis, "θ-QAM: a parametric quadrature amplitude modulation family and its performance in AWGN and fading channels," *Communications, IEEE Transactions on,* vol. 58, pp. 1014-1019, 2010.

[13] J. C. Fricke, M. M. Butt, and P. A. Hoeher, "Quality-oriented adaptive forwarding for wireless relaying," *IEEE Communications Letters,* vol. 12, pp. 200-202, 2008.

[14] S. Bouanen, H. Boujemaa, and W. Ajib, "Threshold-based Adaptive Decode-Amplify-Forward Relaying Protocol for Cooperative Systems," in *IEEE 7th International Wireless Communications and Mobile Computing Conference (IWCMC)*, Canada 2011, pp. 725 - 730

[15] I. Land and P. A. Hoeher, "New results on Monte Carlo bit error simulation based on the a posteriori log-likelihood ratio," in *Proc. 3rd Int. Symp. on Turbo Codes & Rel. Topics (ISTC)*, Brest, France, 2003, pp. 531-534.

[16] A. H. M. Hassan, B. Dai, H. Benxiong, and a. M. A. Iqbal, "Network Coding for Two-Way Relaying over AWGN Channels using Soft Detection " in *Wicom*, Wuhan, 2011.

[17] A. H. M. Hassan, B. Dai, H. Benxiong, and M. A. Iqbal, "Joint design of channel coding and network coding of different channel models for wireless networks," in *IEEE 3rd International Conference on Communication Software and Networks (ICCSN)*, 2011, pp. 46-51.78

**Authors:**

**Ahmed Hassan** received his B.Eng. degree in Electrical Engineering from the Blue Nile University, Sudan, in 2003 and his Master degree in Electronics and Information Engineering from the Sudan University of Science and Technology, Sudan, in 2006. He is currently a Ph.D. candidate at the Department of Electronics and Information Engineering in Huazhong University of Science and Technology, China. His research interests include wireless network coding and cooperative networks. He is an IEEE member.

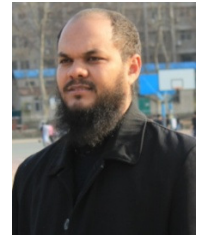

E-mail address: engahmed5577@gmail.com

**Bin Dai** received the B. Eng, the M. Eng degrees and the PhD degree from Huazhong University of Science and Technology of China, P. R. China in 2000, 2002 and 2006, respectively. From 2007 to 2008, he was a Research Fellow at the City University of Hong Kong. He is currently an associate professor at Department of Electronic and Information Engineering, Huazhong University of Science and Technology, P. R. China. His research interests include p2p network, wireless network, network coding, and multicast routing.

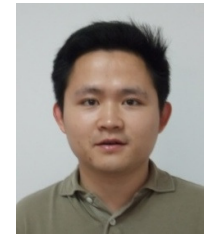

Email address: nease.dai@gmail.com

**Benxiong Huang** received the BS.C in 1987 and PhD in 2003 from Huazhong University of Science and Technology (HUST), P.R China. He is currently professor in the Department of Electronic and Information Engineering at HUST. His research interests include next generation communication system and communication signal processing.

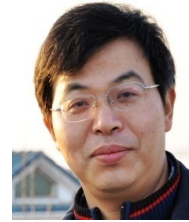

Email address: bxhuang@hust.edu.cn